\documentclass[aps,prb,10pt,twocolumn,amsmath,amssymb]{revtex4-2}

\usepackage{graphicx}
\usepackage{amsmath}
\usepackage{dcolumn}
\usepackage{bm}
\usepackage{xcolor}
\usepackage{graphicx}

\begin{document}

\title{Antiferromagnetism in two-dimensional materials: progress and computational challenges}

\author{Thomas Olsen}

\affiliation{Computational Atomic-Scale Materials Design (CAMD), Department of Physics, Technical University of Denmark, 2800 Kgs. Lyngby, Denmark}
\email{tolsen@fysik.dtu.dk}
%\ead{submissions@iop.org}
%\vspace{10pt}
%\begin{indented}
%\item[]August 2017
%\end{indented}

\begin{abstract}
We present a perspective on the status of antiferromagnetism in two-dimensional (2D) materials. Various types of spin-compensated orders are discussed and include non-collinear order, spin spirals and altermagnetism. Spin-orbit effects ultimately determine, whether compounds exhibit long range order, Kosterlitz-Thouless physics, or multiferroic properties and we discuss the basic magnetic prototypes that may arise in 2D materials depending on the magnetic anisotropy and ordering vector. A summary of 2D antiferromagnets that have been characterized experimentally is provided - with particular emphasis on magnetic anisotropies and Neel temperatures. We then outline the ingredients needed to describe the magnetic properties using density functional theory. In particular, the systematic determination of magnetic ground states from the generalized Bloch theorem and the magnetic force theorem, which may be used to calculate magnetic excitations from the Heisenberg model with parameters determined from first principles. The methods are exemplified by application to the monolayer helimagnet NiBr$_2$. Finally, we present a summary of predicted and prospective 2D antiferromagnets and discuss the challenges associated with the prediction of Néel temperatures from first principles.
\end{abstract}

\maketitle

\section{Introduction}
Magnetic order in bulk three-dimensional materials is traditionally understood in terms of spontaneous symmetry breaking under which the magnetization acquires a definite orientation in space. At the critical temperature ($T_\mathrm{C}$), order is lost due to thermal fluctuations and Weiss mean field theory predicts that $T_\mathrm{C}$ is scales with the energy cost of flipping a single atomic magnetic moment. In two-dimensions (2D), however, the Mermin-Wagner theorem implies that spontaneous symmetry breaking of continuous spin rotational symmetry is not possible \cite{Mermin1966}. The free energy can be lowered by magnetic disorder at any temperature and magnetic order in the traditional sense is therefore not possible at finite temperatures. Nevertheless, relativistic effects give rise to magnetic anisotropy, which introduces a preferred direction relative to the crystal lattice and the symmetry may thus be {\it explicitly} broken, which makes it possible for 2D materials to exhibit magnetic order at finite temperatures. It is important to realize that this mechanism is fundamentally different from that of spontaneously broken symmetry and the critical temperature for disorder is determined by an intricate interplay between magnetic anisotropy and exchange interactions rather than the energy cost of a spin flip.

The current interest in 2D magnetism was largely spurred by the demonstration of ferromagnetic order in a monolayer CrI$_3$ in 2017 \cite{Huang2017}. As anticipated, magnetic order arises due to strong easy-axis (out-of-plane) anisotropy, which breaks the spin rotational symmetry explicitly. Moreover, similar studies on the van der Waals bonded compound CrGeTe$_3$ \cite{Gong2017a} showed ferromagnetic order in thin films down to bilayer structures. However, order is lost in the monolayer limit, which can be assigned to an easy-plane anisotropy that preserves the prerequisites of the Mermin-Wagner theorem. Since these seminal works, much effort has been put into augmenting the pool of known ferromagnetic monolayers, which now includes Fe$_2$GeTe$_3$ \cite{Fei2018}, CrBrS \cite{Lee2021}, MnSe$_2$ \cite{OHara2018} as well as CrCl$_3$ \cite{Bedoya-Pinto2021} and CrBr$_3$ \cite{Zhang2019}. In addition, a large number of magnetic van der Waals bonded materials are composed of ferromagnetic monolayers that may retain the magnetic order upon exfoliation \cite{mcguire2017crystal, Gibertini2019, Jiang2021, Sethulakshmi2019}.

While ferromagnetic order has played a central role in modern technological advances - fx in power generators and digital data storage - the vast majority of current research in atomistic magnetism is focused on antiferromagnetic order. This is largely due to the rich and subtle physics underlying antiferromagnetism, which is - in some regards - still not completely understood. For example, the ground state of the Heisenberg model with antiferromagnetic nearest neighbor exchange strongly depends on the lattice and even for bipartite lattices, the ground state is a correlated state that cannot simply be viewed as a state with anti-aligned spins on the two sub-lattices \cite{Yosida1996}. Moreover, if magnetic frustration is strong, the ground state may become a quantum spin liquid with complete lack of magnetic order \cite{Savary2017}. Finally, antiferomagnetism is often found in close proximity to unconventional superconductivity, and it is believed that the pairing mechanism originates from magnetic correlations in such systems \cite{Stewart2017}. Despite, these fundamentals subtleties, the lack of stray fields in antiferromagnets renders them a promising venue for spintronics applications \cite{Baltz2018}. In this regard, device fabrication typically involves growing heterostructures consisting of layers with different magnetic properties. 2D magnetic materials and their heterostructures thus constitutes an ideal platform that would be able to short-circuit the rather intricate growth process involved in the current device fabrication \cite{Sierra2021}. However, any technological relevance of 2D magnetism relies on magnetic order at room temperature, which has proven exceedingly hard to realize in 2D materials.
\begin{figure*}[tb]
    \centering
    \includegraphics[width=\linewidth]{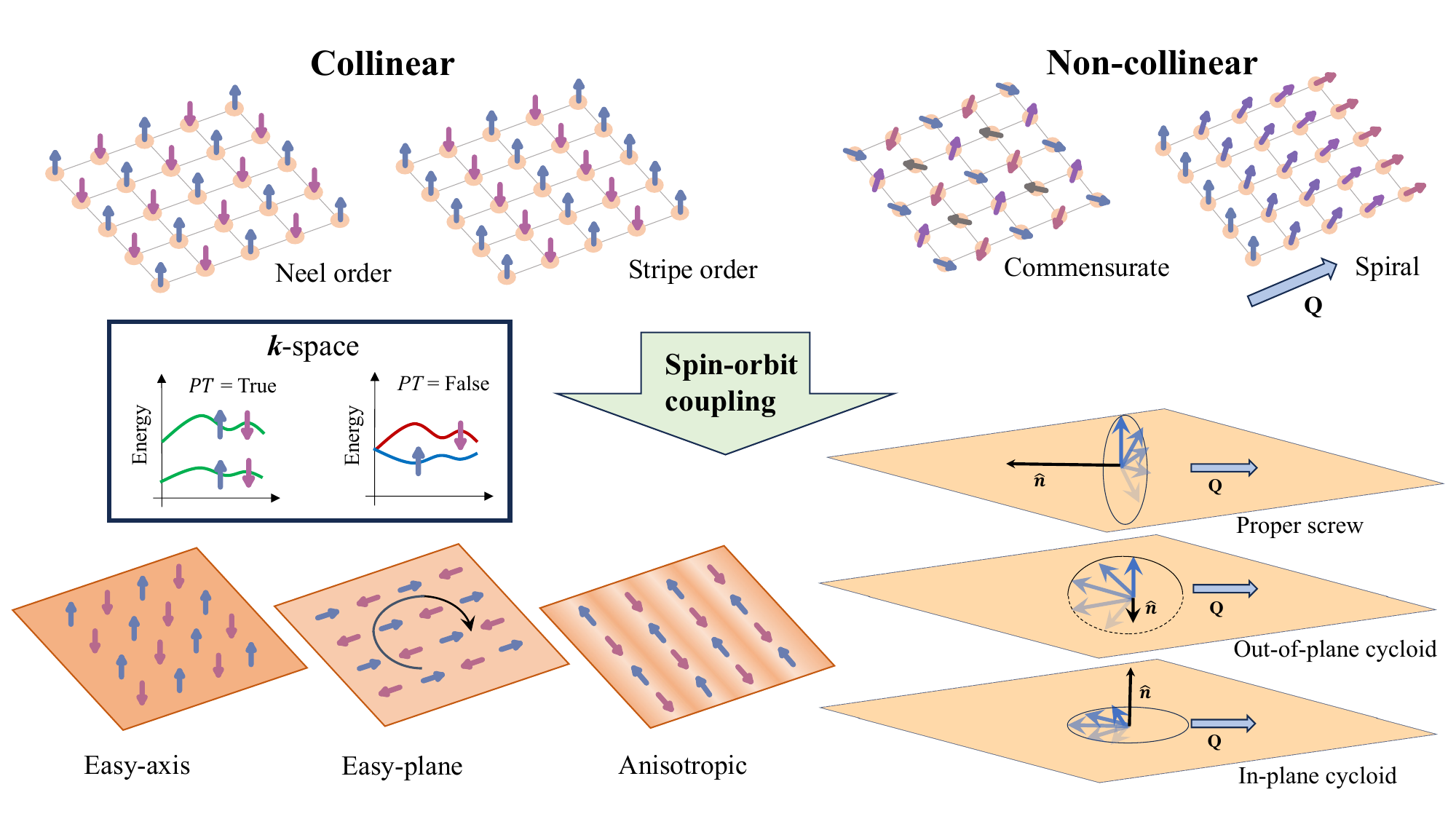}
    \caption{Different types of antiferromagnetism in 2D. At the top we show two types of collinear order as well as examples of commensurate and non-commensurate (spiral) non-collinear order. The lower part shows the effect of spin-orbit coupling exemplified by different limiting prototypes. The box illustrates the fact that collinear order gives rise to spin-splitting in $k$-space in the absence of $PT$ symmetry (altermagnetism).}
    \label{fig:overview}
\end{figure*}

In this perspective we briefly present the current status and computational challenges associated with 2D antiferromagnetism. The focus will be on monolayers. Bilayers, which often exhibit antiferromagnetic interlayer order, will not be discussed. We start with a summary of the prototypical types of antiferromagnetism that may arise in 2D. The most prominent antiferromagnetic monolayers that have been characterized experimentally is discussed individually and compared with their van der Waals bonded bulk parent compounds. We then outline the most commonly applied theoretical approaches to modelling 2D antiferromagnets regarding  both {\it ab initio} approaches to ground state order as well as magnetic excitations and thermodynamic properties extracted from Heisenberg models.

\section{Prototypes of 2D antiferromagnetic order}
There is a multitude of possibilities for antiferromagnetic ordering in a given lattice. A systematic treatment relies on magnetic space groups or magnetic layer groups for 2D materials, but here we will simply focus on particular generic prototypes that covers most known magnetic ordering types in 2D. This classification is much simpler than the corresponding classification in 3D and relies of the unique out-of-plane direction, which is always present in 2D materials. We include both commensurate non-collinear order as well as (possibly incommensurate) spin spiral order and we will thus refer to antiferromagnetic order as any spin-compensated structure. 

In figure \ref{fig:overview} we show examples of 2D antiferromagnetic prototypes. If one neglects spin-orbit coupling, only the relative orientation of spins is meaningful and this may be either collinear or non-collinear. For the collinear case the two most basic types are Néel order and stripe order. The former will comprise the ground state for any bipartite lattice with nearest neighbor isotropic antiferromagnetic exchange and probably constitutes the most intuitive type of antiferromagnetism where any given spin is anti-aligned with all nearest neighbors. Stripe order appears naturally whenever the lattice itself is anisotropic such that ferromagnetic exchange dominates in one direction while antiferromagnetic exchange dominates in an orthogonal direction. It also appears commonly in isotropic systems due to exchange frustration and thus leads to spontaneously broken rotational symmetry. Non-collinear order always arises as a consequence of frustration - typically geometric. For example, the hexagonal lattice with nearest neighbour antiferromagnetic exchange interactions cannot satisfy anti-alignment between all nearest neighbors and will exhibit a planar $120^\circ$ ordered structure as the ground state. Finally, helical order occurs frequently whenever sizeable exchange interactions are not restricted to those between nearest neighbours. It is, in fact, straightforward to show that the energy of the isotropic classical Heisenberg model on a Bravais lattice is minimized at the wave vector $\mathbf{Q}$ that maximizes the Fourier transform of the exchange interactions. In this regard, collinear ferromagnetic and antiferromagnetic order emerges as limiting cases where $\mathbf{Q}$ is situated at certain high symmetry points in the Brillouin zone and only appear rather commonly because magnetic interactions are often dominated by nearest neighbor exchange. 

Spin-orbit coupling (SOC) gives rise to magnetic anisotropy, which determines the orientation of the magnetic moments with respect to the lattice. Most of the 2D materials that have been studied intensively exhibit at least three-fold rotational symmetry in the plane, which renders the in-plane components of second rank tensors isotropic. For collinear order this implies a natural distinction between easy-axis or easy-plane materials and as a consequence of the Mermin-Wagner theorem only the former can exhibit strict long range order. It is, however, well-known that the XY-model (spins confined to a plane) undergoes a Kosterlitz-Thouless transition \cite{Thouless1973} below which quasi long-range order appear and this may be indistinguishable from proper long range order for most practical purposes \cite{Holdsworth1994} although the critical behaviour will be different. Nevertheless, the extend to which easy-plane 2D collinear magnets mimics the physics of the XY model is not clear and it is possible that such behaviour is only observable in cases where the magnetic anisotropy is exceedingly strong. In this regard, it should also be noted that the assumption of in-plane isotropy is only valid in models that are quadratic in the spin degrees of freedom. Higher order interactions may break the in-plane rotational degree of freedom \cite{sødequist2023magnetic} and thus give rise to proper long range order. Although isotropic monolayers appear to be more common, anisotropic monolayers have been realized (CrBrS comprises a ferromagnetic example that has recently been scrutinized \cite{Zhang2019}) and such cases naturally exhibit triaxial anisotropy, which is always accompanied by magnetic order if the temperature becomes sufficiently low.

The inset of figure \ref{fig:overview} shows a sketch of an antiferromagnetic band structure. Most known antiferromagnets exhibit a combination of inversion ($P$) and time-reversal ($T$) symmetry, which enforces spin-degeneracy of the bands \cite{Yuan2020}. However, in the absence of this symmetry the bands become non-degenerate and for metals, in particular, this could constitute a promising platform for non-ferromagnetic spintronics applications \cite{Smejkal2022}. Such spin-splitting will even occur without spin-orbit coupling in materials belonging to the type III magnetic space groups, (denoted altermagnets) and the spin-splitting in these materials can thus become much larger than typical spin-orbit energies.

Planar spin spirals (including non-collinear commensurate order) allow for a similar classification in terms of 2D prototypes \cite{Sødequist2023}. SOC determines the orientation of the spiral plane with respect to the lattice and the limiting cases are the proper helix (spiral plane orthogonal to $\mathbf{Q}$), the in-plane cycloid (spiral plane coincides with atomic plane) and the out-of-plane cycloid (spiral plane spanned by $\mathbf{Q}$ and normal vector to atomic plane). These prototypes are sketched in figure \ref{fig:overview}. An interesting aspect of spiral order is the fact that it typically breaks inherent symmetries of the lattice and implies that the magnetic order may induce ferroelectric order. For a spiral plane with normal vector $\mathbf{\hat n}$ such type II multiferroics have been predicted to yield a polarization $\mathbf{P}\propto\mathbf{Q}\times\mathbf{\hat n}$, which thus excludes the case of proper helices \cite{Katsura2005}. Recent first principles computations has, however, cast doubt on the validity of this prediction \cite{Xiang2011, Song2022}, but the orientation of the spiral plane does seem to play a dominant role for the magnitude and orientation of the induced polarization \cite{Sødequist2023}. We finally note that spiral order may be induced by either isotropic exchange or Dzyaloshinskii-Moriya (DM) interactions in materials lacking inversion symmetry. In may in fact be shown that inclusion of DM interactions in the isotropic classical Heisenberg model always yields a spiral ground state with the spiral plane orthogonal to the DM vector \cite{Schweflinghaus2016}. In the general case, this only happens if the DM energy is larger than the anisotropy lost by the spiral (compared to a collinear state) \cite{sødequist2023magnetic}.

In the presence of higher order spin interactions, the ground state may not be described by a planar spin spiral. In particular, for bilinear spin models a spiral at a particular $\mathbf{Q}$ will be degenerate with all wave vectors related by symmetry, but if higher order interactions are included it becomes favorable to form linear combinations of these, which may yield either non-coplanar order or collinear structure that cannot be described by a single ordering vector \cite{Gutzeit2022}. Non-coplanar structures exhibit a range of intriguing properties such as non-relativistic anomalous Hall conductivity and orbital magnetization \cite{Hanke2016} and the fact that these quantities are not driven by spin-orbit coupling implies that they may become much larger compared to typical values in collinear systems. In the presence of DM interactions, such materials may form skyrmion lattices \cite{Muhlbauer2009}, which has been observed in thin films \cite{Yu2011, Heinze2011} and heterostructures \cite{Soumyanarayanan2016a} but not yet (to our knowledge) in a proper 2D material.

\section{Observations of 2D antiferromagnetism}
In nearly 80 years neutron diffraction has been the major work horse for extracting magnetic structures of bulk materials \cite{WORACEK2018141} and it was applied to explicitly demonstrate the existence of antiferromagnetic order in MnO \cite{PhysRev.76.1256.2}. However, due to the reduced sample volume it is presently not possible to perform neutron diffraction on 2D materials, and the demonstration of magnetic order in ferromagnets has instead been achieved by use of magneto-optical Kerr effect \cite{Huang2017}, circularly polarized photoluminescence \cite{Seyler2017}, reflective magnetic circular dichroism \cite{Cenker2021} or the anomalous Hall effect (for metals) \cite{Fei2018}. In antiferromagnets, non of these methods can be applied and the extraction of magnetic order is more difficult, but may be accomplished using either Raman spectroscopy \cite{Lee2016, Kim2019}, spin-polarized scanning tunnel microscopy \cite{Chang2013a, Xian2022} or second harmonic generation \cite{MnPSe3}.

\begin{figure*}[tb]
    \centering
    \includegraphics[width=\linewidth]{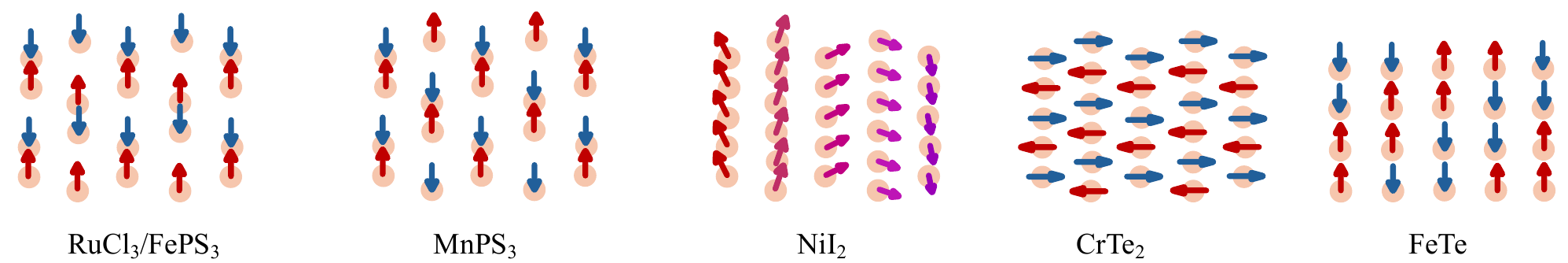}
    \caption{Antiferromagnetic order in selected 2D materials. Only the magnetic atoms are shown. NiPS$_3$ has the same order as FePS$_3$ and RuCl$_3$, but with an easy-plane instead of an easy-axis.}
    \label{fig:spin}
\end{figure*}

Below we briefly summarize some of the experimental progress that has been achieved over the past eight years in scrutinizing antiferromagnetic order in 2D materials. Table \ref{tab:antiferromagnets} summarizes some of the key quantities of the reviewed materials and figure \ref{fig:spin} illustrates the magnetic structures.
\begin{table}[t]
\begin{tabular}{l|c|c|c|c|c}
                 & Space group  & Order & Alignment  & T$_\mathrm{N}^\mathrm{bulk}$ & T$_\mathrm{N}^\mathrm{2D}$  \\
                 \hline
FePS$_3$ \cite{Lee2016}               & $P\bar31m$   & Zigzag  & $\hat{z}$ & 123 &  118 \\
MnPS$_3$ \cite{Long2020}              & $P\bar31m$   & Néel    & $\hat{z}$ & 78  &  78  \\
MnPSe$_3$ \cite{MnPSe3}               & $P\bar31m$   & Néel    & $\hat{x}$ & 74  &  40  \\
NiPS$_3$ \cite{Kim2019b}              & $P\bar31m$   & Zigzag  & $\hat{x}$ & 155 &   -  \\
RuCl$_3$ \cite{Yang2023}              & $P\bar31m$   & Zigzag  & $\hat{z}$ &  7  & 8    \\
NiI$_2$  \cite{Song2022}              & $P\bar3m1$   & Spiral  & -  & 75 & 21 \\
CrTe$_2$ \cite{Xian2022}              & $P2/c$       & Zigzag  & $\hat{x}$ & 310 & 320 \\
FeTe     \cite{Kang2020}              & $P4/nmm$     & Stripe  & $\hat{x}$ & 70 & $< 45$
\end{tabular}
\caption{List of exfoliated antiferromagnetic monolayers. The alignment of magnetic moments is denoted by either $\hat{z}$ (out-of-plane) or $\hat{x}$ (in-plane).}
\label{tab:antiferromagnets}
\end{table}

\subsection{MPS$_3$}
The transition metal phosphorus chalcogenides (MnPS3 with M=Mn,Fe,Co,Ni) constitutes an versatile class of van der Waals bonded antiferromagnets \cite{Joy1992, Wang2018} that have been  characterized in detail by neutron scattering - both in terms of ground state order \cite{Wildes2015} and magnetic excitations \cite{Wildes1998, Lancon2018}. The magnetic atoms in a single layer forms a honeycomb lattice with different types of magnetic order depending on the transition metal. The Fe, Ni and Mn compounds have been studied most and MnPS$_3$ exhibits Néel order while FePS$_3$ and NiPS$_3$ have a zigzag ordered ground states. The Néel temperatures are 78, 123 and 155 K respectively \cite{Joy1992}. In addition, while the Fe and Mn compounds exhibit out-of-plane easy-axes the Ni compound has an easy-plane. Despite having similar crystal fields and ligands these three materials show distinct types of magnetic order, which makes them highly suited for fundamental scrutiny in microscopic magnetic interactions. In addition, the magnetic anisotropy energies differ by more than three orders of magnitude \cite{Lancon2018} with Fe having the largest and Mn having the smallest anisotropy energy. The spin states of these compounds are $S=2$, $S=1$ and $S=5/2$ respectively and the low anisotropy of the Mn compound is easily rationalized from the approximate $L_z=0$ implied by the spin state. The Fe compound has an unusually high anisotropy for a $3d$-metal, which seems to defy predictions of first principles calculations \cite{Olsen2021}. Nevertheless, in Ref. \cite{Kim2021} it was shown that applying Hubbard corrections in conjunction with self-consistent SOC yields an unprecedented large orbital moment on the Fe atoms ($\sim1\;\mu\mathrm{B}$), which is pinned to the lattice rather than the spin. This large orbital moment and the associated enormous magnetic anisotropy was recently explicitly verified experimentally \cite{Lee2023}. The crucial importance of including self-consistent SOC, which is usually regarded as a small perturbation, highlights the subtleties and caveats involved in first principles simulations of magnetism. In addition to the Mn, Fe and Ni compounds, magnetic order has also been observed in CoPS$_3$ \cite{Wildes2017}, which shows the same type of order as NiPS$_3$ with a Néel temperature of 120 K. Finally, NiPSe$_3$ and FePSe$_3$ have been shown to exhibit similar magnetic ordering \cite{FERLONI1989197}, but the Se compounds seem to have been studied somewhat less compared to the S compounds.

The Ni, Fe and Mn structures have all been exfoliated to the monolayer limit and the situation regarding magnetic order in the atomically thin limit is rather interesting. Magnetic order in the Fe compound was demonstrated by Raman spectroscopy in 2016 \cite{Lee2016} and thus rightfully constitutes the first report of 2D magnetism - preceeding the demonstration of ferromagnetism in monolayer CrI$_3$. Magnetic order in FePS$_3$ appears to be driven by the Ising type (strong easy-axis anisotropy) magnetic order and the Néel temperature for the monolayer was reported to be 118 K - very close to the 123 K of bulk. 

In MnPS$_3$, the weak anisotropy renders the bulk materials nearly isotropic (albeit with a weak out-of-plane easy-axis) and it is {\it a priori} far from clear whether order will persist in the monolayer limit. In particular, Raman measurements have suggested that monolayer MnPS$_3$ is non-magnetic \cite{Kim2019a}. On the other hand, in Ref. \cite{Long2020} the authors exploited the spin-flop transition associated with the weak anisotropy and tracked it by tunneling magnetoresistance measurements as the sample thickness was reduced. They concluded that magnetic order persists in MnPS$_3$ all the way to the monolayer limit without significant change in the Néel temperature. The theoretical understanding of magnetism is MnPS$_3$ seems to hinge on the effects of magnetic dipole interactions \cite{Wildes1998}. While these are usually weak (on the order of $\mu$eV) they play a dominant role in MnPS$_3$ due to the $S=5/2$ state of the Mn atoms, which renders exchange anisotropy negligible \cite{Kim2021}. Again, this puts severe demands on any first principles attempt of calculating fundamental anisotropy parameters, since it requires computations with an accuracy well below 1 meV per magnetic atom. 

The case of NiPS$_3$ has been shown to exhibit a magnetic Raman peak that could be observed for samples thinned all the way to the bilayer limit - but not for the monolayer \cite{Kim2019b}. Due to the easy-plane anisotropy of NiPS$_3$ it is natural to interpret this as the working principle of the Mermin-Wagner theorem that prohibits order in the strict 2D limit (although the possibility of Kosterlitz-Thouless physics remains). Nevertheless, there has been cast some doubt on this conclusion due to the enhanced symmetry in the monolayer limit, which it expected to make the Raman signal disappear if multiple domains are present \cite{Kim2021}. It remains to be seen whether or not monolayer NiPS$_3$ exhibits proper long range order despite the presence of an easy-plane.

Finally, the sister compounds FePSe$_3$ and MnPSe$_3$ have been studied by neutron diffraction and have Néel temperatures of 119 and 74 K respectively \cite{WIEDENMANN19811067}. In addition MnPSe$_3$ has recently been exfoliated and shown to have strong easy-plane (in sharp contrast to MnPS$_3$) and a Néel temperature of 40 K \cite{MnPSe3}.

\subsection{NiI$_2$}
NiI$_2$ is a well known van der Waals bonded compound that crystallizes in the $R\bar3m$ space group (ABC stacking) whereas the individual layers are of the 1T type (space group $P\bar3m1$) \cite{mcguire2017crystal}. The magnetic order of the bulk is helimagnetic \cite{KUINDERSMA1981231}, with a spiral vector of $\mathbf{Q}=(0.138, 0, 1.457)$ (in reciprocal lattice units) that roughly corresponds to and in-plane spiral modulation while the spin is flipped between layers. The spiral plane is tilted by $55^\circ$ relative to the atomic planes, which renders the magnetic structure close to a 3D proper helix with a $\mathbf{Q}$-vector tilted by $35^\circ$ relative to the atomic planes.

In Ref. \cite{Song2022} it was demonstrated that spiral order is maintained in monolayer NiI$_2$ below 21 K. This was accomplished by angular-dependent linear dichroism measurements that showed that rotational symmetry was broken below the critical temperature. This is a hallmark of spiral order, which thus renders the structure ferroelectric and NiI$_2$ comprises the first experimental demonstration of a type II multiferroic 2D material.  Recently, the magnetic order in monolayer NiI$_2$ was further elucidated by scanning tunneling microscopy, which confirmed the spiral order and determined the 2D spiral vector to be $\mathbf{Q}=(0.069, 0.041)$ \cite{amini2023atomicscale}. However, the orientation of the spiral plane as well as the magnitude of spontaneous polarization has not been measured yet.

Separate groups have shown that DFT calculations of NiI$_2$ predict a ground state with spiral order in agreement with experiments as well as an induced in-plane polarization, but there seems to be some disagreement of the orientation of the spiral plane as well as the magnitude of polarization. In Ref. \cite{Song2022} the exchange constants were calculated and used in conjunction with classical Monte Carlo simulations to predict a critical temperature in good agreement with the measurements while the spiral order was predicted to be a 2D proper helix. In Ref. \cite{Fumega2022} an in-plane cycloid was assumed, which yielded a polarization of 0.5 pC/m (from an inverse DM model), whereas Ref. \cite{Sødequist2023} predicted a spiral plane tilted by $64^\circ$ with respect to the atomic plane and a polarization of 1.8 pC/m (by explicit first principles calculations). While the latter result seems to be in good agreement with the bulk spiral plane orientation, the measured polarization in bulk implies that each monolayer contributes a mere 0.08 pC/m \cite{mcguire2017crystal}. Nevertheless, this value may simply represent a lower bound due to the presence of multiple domains in the samples.

In bulk materials, spiral order appears to be rather common and a compilation of all experimentally known van der Waals bonded transition metal dihalides has shown that 5 out of 14 compounds exhibit in-plane spiral order, while two compounds show non-collinear commensurate order. In Ref. \cite{sødequist2023magnetic}, it was found that 51 out of 164 simple magnetic 2D materials from the Computatational 2D Materials Database (C2DB) \cite{Haastrup2018, Gjerding2021} exhibited a ground state with incommensurate spiral order. However, it should be stressed that accurate predictions of spiral plane orientations and induced polarizations constitutes a major challenge for DFT due to the intricate spin-orbit effects that are involved as well as the (possibly) strong correlation often present in such compounds.

\subsection{CrTe$_2$}
Bulk CrTe$_2$ has been synthesised in the 1T-phase (AA stacking with space group $P\bar3m1$) and exhibits itinerant (half-metallic) ferromagnetic order below 310 K \cite{Freitas_2015} with easy-plane anisotropy. Subsequently, a monolayer of 1T-CrTe$_2$ was grown on bilayer graphene and spin-polarised STM showed that the monolayer retains the half-metallic behavior and becomes antiferromagnetic with zigzag order \cite{Xian2022}. In addition, the structure undergoes a structural distortion that lowers the symmetry to $P2/c$ (or so it appears) that breaks rotational symmetry. DFT calculations imply that the structural change is, however, not driven by the magnetic order since it happens in the ferromagnetic state as well \cite{Haastrup2018,Manti2023}. The broken symmetry allows for an easy axis in the monolayer and DFT calculations was applied to predict an orientation which is tilted off the atomic plane. Second harmonic generation was applied to demonstrate that magnetic order persists up to $\sim 320$ K.

The emergence of itinerant antiferromagnetism at room temperature in a 2D material is highly intriguing, but several open questions needs to be addressed in order to explain these observations properly. First of all, the easy-plane of the bulk compound implies that magnetic order hinges crucially on interlayer exchange interactions, but since these are typically rather weak in van der Waals bonded materials it is difficult to rationalise the high Curie temperature in bulk. Second, the transition to antiferromagnetic order in the monolayer limit signals a sign change in the intralayer magnetic interactions, which implies rather weak exchange. Moreover, the magnetic anisotropy is similar in magnitude to the case of CrI$_3$ and it is hard to understand the high Néel temperature of the monolayer in simple terms of the broken rotational symmetry of the atomic plane. It is likely that the itinerant nature of the compound plays a crucial role in this regard, but also makes it much more challenging to handle theoretically. In particular, it is not obvious that the standard model of magnetism - the Heisenberg model - is applicable for metals and the adiabatic assumption \cite{Halilov1998} often applied in DFT calculations of magnetic interactions may break down.

\subsection{RuCl$_3$}
Bulk RuCl$_3$ (often referred to as $\alpha$-RuCl$_3$) crytallizes in the $C2/m$ monoclinic space group at low temperatures \cite{Cao2016}. The structure is composed of AB stacked monolayers each forming a honeycomb Ru lattice sandwiched between Cl atom (similar to CrI$_3$) with space group $P\bar31m$. Neutron diffraction has determined the most likely magnetic order to be a collinear antiferromagnetic zigzag structure similar to the case of FePS$_3$ with the easy axis tilted by $35^\circ$ from the atomic plane and a Néel temperature of 7 K. The compound has received considerable amount of attention due to to strong spin-orbit effects that imply an effective spin-orbit entangled $J=1/2$ system. In addition, evidence of a continuum of magnetic excitations suggests that the ordered ground state exhibits Kitaev physics and resides in close proximity to a quantum spin liquid state \cite{Banerjee2016, Banerjee2017}.

Monolayers of RuCl$_3$ have been exfoliated and maintain the intralayer antiferromagnetic order of below 8 K while the magnetic anisotropy rotates to an out-of-plane easy-axis \cite{Yang2023}. In addition, it has been shown that the magnetic order can be changed to ferromagnetic by gating or optical pumping \cite{Tian2019}. Computational approaches for describing the magnetic interactions has proven rather challenging due to inherent Mott insulating behaviour. In particular, DFT approaches requires Hubbard correction to obtain a band gap, but the magnetic interactions are rather sensitive to these and some DFT+U approaches have even predicted a ferromagnetic ground state \cite{Sarikurt2018}.

\subsection{FeSe/FeTe}
FeTe and FeSe have been studied intensely over the past decade due to the emergence of unconventional superconductivity at low temperatures \cite{Si2016}. The pairing mechanism is believed to be mediated by magnetic fluctuations and the superconducting phase is found in close proximity to antiferromagnetic order. FeTe, however, only becomes superconducting upon doping with S or Se atoms, but the pristine materials exhibits striped AFM order with $\mathbf{Q}=(1/2,1/2,0)$ below 70 K. On the other hand, pristine FeSe becomes superconducting below 8 K , but does not exhibit magnetic order \cite{Kreisel2020}. Nevertheless, despite the lack of magnetic order. neutron scattering has shown that FeSe exhibits spin fluctuations consistent with $\mathbf{Q}=(1/2,0,0)$ order and DFT calculations have shown that different magnetic configurations result in very similar energies \cite{Wang2015}. The lack of magnetic order in FeSe has been explained by the presence of bi-quadratic spin interactions that destabilize the ground state \cite{Kreisel2020}, but situates the compound in close proximity to an ordered state.

Monolayers of FeTe has been grown on SrTiO$_3$ \cite{PhysRevB.91.220503} and found to lack superconducting order, while Se doped monolayers exhibits superconductivity \cite{PhysRevB.100.155134,PhysRevB.108.214514}. While the magnetic properties of the FeTe monolayers were not investigated, the lack/presence of a superconducting gap without/with Se doping mimics the situation in bulk FeTe/FeSe and was assigned to the presence of antiferromagnetic order in monolayer FeTe. In addition, monolayers of FeTe has been grown on the topological insulator Bi$_2$Te$_3$ where superconducting order coexist with AFM order up to 6 K. The magnetic order does, however, not vanish above the superconducting transition and the Néel temperature was found to be larger than 10 K. Finally, in Ref. \cite{Kang2020}, thin films of FeTe was grown on Si/SiO$_2$ substrates and the Néel temperature was shown to decrease from 70 K to 45 K for 5 nm samples.

\section{Computational approaches to antiferromagnetic monolayers}
For insulators, the magnetic properties are typically well represented by the Heisenberg model \cite{Yosida1996} and the quest for a first principles description of magnetism then becomes a matter of reliable predictions of parameters in the model. However, solving the Heisenberg model is, in general, a non-trivial task and thermal properties are usually extracted from either classical Monte Carlo (MC) simulations or the random phase approximation (RPA) \cite{tyablikov1959}. The performance of both methods becomes questionable for elevated temperatures (probably for all temperatures in the case of classical MC) and the accuracy of predicted critical temperatures is often rather dubious. The exchange constants can be used to calculate magnon dispersion relations within the model and thus validate first principles approaches for obtaining accurate parameters in cases where the magnon dispersion has been measured. For most magnetic materials, the magnon dispersion is, however, not known and any assessment of first principles approaches describing magnetic properties has to rely on comparison of thermal properties - for example the critical temperature for magnetic order. But since it is not clear how to obtain accurate estimates for thermal properties from the Heisenberg model (at least when approaching the critical temperature), deviations between theory and experiments may originate from both inaccurate treatment of the Heisenberg model and from inaccurate first principles calculations of exchange parameters (for example from DFT). As an example, the exchange constants of ferromagnetic 2D CrI$_3$ have been calculated by several different groups with varying results depending on methodology and exchange-correlation functional. The common approach of PBE+U yields rather large exchange constants that produce an optical magnon branch that is overestimated by more than 50 {\%} \cite{Olsen2019}. Classical MC simulations with these exchange constants, however, yield a Curie temperature of 45 K in agreement with experiments - probably because of rather severe underestimation of the critical temperatures in the classical MC approach \cite{Torelli2018}. Moreover, the sizable magnon gap at the Dirac point in bulk CrI$_3$ has been assigned to both DM interactions \cite{Chen2018a} and Kitaev interactions \cite{Lee2020} in experiments and it is currently not clear whether DFT is able to resolve this controversy.

Magnetic excitations can also be calculated directly from first principles methods such as time-dependent density functional theory (TDDFT) \cite{Buczek2011, Cao2018, Skovhus2021} and many-body perturbation theory (MBPT) \cite{Sasoglu2010, Muller2016, Olsen2021a}. Although either approach is rather demanding in terms of computational load, they have the advantage of being directly comparable to experiments without having to justify the validity of mapping to a site-based model. It is, however, not possible to include thermal effects with these methods. Dynamical mean field theory (DMFT) \cite{Georges1996} is the only approach that (in principle) is able to handle finite temperatures as well as strong correlation, but the accuracy may be somewhat limited by the local approximation for correlation effects \cite{Lichtenstein2001}. 

For metals, the situation becomes even more difficult because it is unclear whether or not it is sensible to treat magnetic properties in terms of Heisenberg models with localized spins. While alternative approaches for modelling itinerant magnetism have been proposed \cite{Moriya1991}, the methods have not gained widespread popularity and the general applicability for first principles approaches is perhaps questionable. Alternatively, one may apply an adiabatic approximation for magnons, which restores a classical Heisenberg description of the problem \cite{Halilov1998}, but key itinerant effects such as Landau damping are not captured by such approaches

Below we start by summarizing the central ingredients of the Heisenberg model along with the most common methods for calculating magnon energies and thermal properties. We then outline the DFT-based approaches for calculating magnetic ground states and Heisenberg parameters, but will refrain from discussing TDDFT, MBPT and DMFT further.

\subsection{Heisenberg models}
The Heisenberg model with bilinear spin interactions may be written as
\begin{align}
H=-\frac{1}{2}\sum_{aibj\alpha\beta}S_{ai}^\alpha\mathcal{J}^{\alpha\beta}_{aibj}S_{bj}^\beta-\sum_{ai\alpha\beta}S^\alpha_{ai}\mathcal{A}^{\alpha\beta}_{a}S^\beta_{ai},
\end{align}
where $S_{ai}^\alpha$ denotes the $\alpha$-component of the spin operator for site $a$ in unit cell $i$. $\mathcal{J}^{\alpha\beta}_{aibj}$ is the exchange interaction tensor and $\mathcal{A}^{\alpha\beta}_{a}$ is the symmetric single-ion anisotropy tensor for site $a$.  It is often convenient to decompose the exchange tensor into isotropic, antisymmetric and symmetric traceless parts:
\begin{align}
J_{aibj}&=\mathrm{Tr}\mathcal{J}^{\alpha\beta}_{aibj}/3,\\
D_{aibj}^{\alpha\beta}&=\frac{1}{2}\Big[\mathcal{J}^{\alpha\beta}_{aibj}-\mathcal{J}^{\beta\alpha}_{aibj}\Big],\\
K_{aibj}^{\alpha\beta}&=\frac{1}{2}\Big[\mathcal{J}^{\alpha\beta}_{aibj}+\mathcal{J}^{\beta\alpha}_{aibj}\Big]-J_{aibj}\delta^{\alpha\beta}.
\end{align}
The non-isotropic part arises as a consequence of spin-orbit coupling. Its antisymmetric part can be written as an axial vector known as the DM vector and gives rise to chiral magnetic effects - for example skyrmions, spin canting and spin spirals. The symmetric traceless part can be related to the Kitaev interactions, which have been shown (neglecting all other terms) to yield a quantum spin liquid ground state on the honeycomb lattice \cite{Savary2017}. There has been intense search for materials exhibiting the Kitaev quantum spin liquid state and RuCl$_3$ has been a major contender for a several years \cite{Banerjee2016, Banerjee2017}.

While the anisotropic exchange interactions give rise to rich magnetic properties, even the correlations of the isotropic model is not very well understood. In the isotropic case the fully polarized state is always an eigenstate, but it only comprises the ground state for ferromagnets. It is straightforward to verify that a state with alternating polarization (maximal or minimal eigenvalues for $S^z$ at all sites) is not an eigenstate and the antiferromagnetic ground state is only known for $S=1/2$ in 1D where the model can be solved by the Bethe Ansatz \cite{Karabach1997}. In that case the ground state is not ordered and the magnetic excitations form a continuum similar to the case of the quantum spin liquids. In higher dimensions, the ground state is ordered and may be approximated by the non-interacting magnon state, which can be shown to have a lower energy than the classical minimum \cite{Yosida1996}.

Although the quantum mechanical ground state of the Heisenberg model is not known in general, it is straightforward to show that the {\it classical} ground state of the isotropic Heisenberg model on a Bravais lattice can be written as a planar spiral of the form
\begin{align}\label{eq:spiral_state}
\mathbf{S}_{i}=R_i(\mathbf{Q})
\begin{pmatrix}
0 \\
0 \\
S,
\end{pmatrix}
\end{align}
where $S$ is the magnitude of the spin (maximal eigenvalue of $S^z$). $R_i(\mathbf{Q})$ is a rotation matrix that rotates the spin in the spiral plane by an angle $\mathbf{Q}\cdot\mathbf{R}_i$ where $\mathbf{R}_i$ is the lattice vector corresponding to unit cell $i$ and the ordering vector $\mathbf{Q}$ is uniquely determined by the exchange constants $J_{ij}$. This includes ferromagnetic order with $\mathbf{Q}=(0, 0, 0)$ and various types of collinear antiferromagnetic order where some of the components of $\mathbf{Q}$ are 1/2 (and the rest vanishes). Although, this is only true in the classical limit, the spiral ansatz \eqref{eq:spiral_state} is widely applied to determine possible magnetic ground states and it comprises the starting point for linear spin-wave theory described below. In the isotropic case the orientation of the spiral plane is not determined by the model, but the ansatz also remains valid upon inclusion of DM interactions, which will serve to determine the orientation of the spiral plane with respect to the DM vectors \cite{Schweflinghaus2016}.

If higher order spin interactions are included in the model the classical ground state is no longer described by a spin spiral. Because of time-reversal symmetry it is not possible to have three-spin interactions and the simplest form of higher order interactions is the isotropic bi-quadratic term
\begin{align}\label{eq:biquadratic}
\Delta H^\mathrm{bq}=-\sum_{aibj}B_{aibj}(\mathbf{S}_{ai}\cdot\mathbf{S}_{bj})^2.
\end{align}
In addition to this it is possible to add various three-site and four-site anisotropic four-spin interactions \cite{Hoffmann2020, Gutzeit2021}. The effect of these terms is a coupling between degenerate spiral vectors, which leads to linear combinations of these that can lower the energy. The ground state is then typically described by a multi-$q$ state that can be either collinear and may be described by a spiral ansatz in a super cell (for example a double striped antiferromagnetic Bravais lattice) or a non-coplanar state (for example skyrmions) \cite{Gutzeit2022}. In addition, bi-quadratic spin interactions have been argued to drive the destabilization of magnetic order in FeSe \cite{Kreisel2020}.

Another effect that may cause anisotropy, is the magnetic dipole-dipole interactions. The classical expression for the energy is
\begin{align}\label{eq:dipole}
\Delta H^\mathrm{md}=-\frac{\mu_0\hbar^2}{8\pi}\sum_{aibj}\frac{\gamma_a\gamma_b}{r_{aibj}^3}\bigg[&3(\mathbf{S}_{ai}\cdot\mathbf{\hat r}_{aibj})(\mathbf{S}_{bj}\cdot\mathbf{\hat r}_{aibj})\\
&-\mathbf{S}_{ai}\cdot\mathbf{S}_{bj}\bigg],\notag
\end{align}
where $\mu_0$ is the vacuum permeability, $\gamma_a$ is the gyromagnetic ratio for site $a$, $\mathbf{\hat r}_{aibj}$ is the unit vector connecting site $ai$ and $bj$ and $r_{aibj}$ is the distance between the two sites. Similar to anisotropic exchange interactions, the dipole-dipole interaction originates from relativistic effects, but in contrast to SOC, it does not appear at the single-particle level and will thus typically not be accounted for by DFT (unless included explicitly in the exchange-correlation functional). For 2D materials, this term will favour in-plane order for ferromagnets and out-of-plane order for collinear antiferromagnets. For bulk materials the dipole-dipole interactions are, however, typically ignored since the interaction energy is on the order of $\mu$eV, whereas the exchange interactions are of the order meV. Nevertheless, in compounds with weak SOC this term may have a decisive effect on the order and for MnPS$_3$ it has been shown that inclusion of dipole-dipole interactions yields a prediction of easy-axis order, whereas SOC alone predicts easy-plane order \cite{Kim2021}. The dipole-dipole induced easy-axis in bulk MnPS$_3$ is in agreement with experiments.

In 2D, anisotropy is required for magnetic order at finite temperatures and an often applied minimalistic model is to include single-ion anisotropy in addition to isotropic exchange. Diagonalizing $\mathcal{A}^{\alpha\beta}$, may either yield two (uniaxial) or three (triaxial) distinct eigenvalues. In the uniaxial case the distinct eigenvalue will correspond to the out-of-plane direction and whether this is the smallest or largest eigenvalue determines if the material has an easy-plane or easy-axis. In the presence of three-fold (or higher) rotational symmetry, magnetic isotropy of the atomic plane is enforced and if this is the preferred direction the system cannot order by the Mermin-Wagner theorem. This statement, however, relies on the bilinear model and in reality there will always be some amount of in-plane anisotropy due to higher order spin interactions. In principle it is therefore always possible for 2D magnets to order - regardless of the presence of symmetries enforcing easy-plane magnetic order - but the ordering temperature may become rather low in such cases.

The classical ground state of the Heisenberg model serves as the starting point for calculating magnon energies within linear spin-wave theory. For a general spiral state described by Eq. \eqref{eq:spiral_state} the magnon dispersion can be obtained by performing a Holstein-Primakoff transformation in a co-rotating frame \cite{Toth2015}. Here we just state the result for a single site in the unit cell for a spiral plane containing the $z$-axis. The basic ingredient is the dynamical matrix
\begin{align}\label{eq:dyn_matrix}
h(\mathbf{q})=
-\frac{S}{2}\begin{pmatrix}
A_{\mathbf{q}}-C   & B_{\mathbf{q}}\\    
B^*_{\mathbf{q}} & A^*_{-\mathbf{q}}-C
\end{pmatrix}
\end{align}
with
\begin{align}
&A_{\mathbf{q}}=\mathcal{J}^{xx}_\mathbf{q}+\mathcal{J}^{yy}_\mathbf{q} - i(\mathcal{J}^{xy}_\mathbf{q}-\mathcal{J}^{yx}_\mathbf{q})\\
&B_{\mathbf{q}}=\mathcal{J}^{xx}_\mathbf{q}-\mathcal{J}^{yy}_\mathbf{q} + i(\mathcal{J}^{xy}_\mathbf{q}+\mathcal{J}^{yx}_\mathbf{q})\\
&C = 2\mathcal{J}^{zz}_\mathbf{0}
\end{align}
where
\begin{align}
\mathcal{J}_\mathbf{q}=\sum_i\mathcal{J}_{0i}R_i(\mathbf{Q})e^{i\mathbf{q}\cdot\mathbf{R}_i}
\end{align}
and $R_i(\mathbf{Q})$ is the rotation matrix that rotates the $z$-axis into the spin direction of unit cell $i$. Single-ion anisotropy may be included by defining $\mathcal{J}_{00}\equiv2\mathcal{A}$. The magnon energy at a given $\mathbf{q}$ can then be found as the positive eigenvalues of the matrix $K_\mathbf{q}gK^\dag_\mathbf{q}$ where $K_\mathbf{q}$  is obtained from a Cholesky decomposition of $h(\mathbf{q})=K^\dag_\mathbf{q}K_\mathbf{q}$ and
\begin{align}\label{eq:g}
g=
\begin{pmatrix}
1  & 0 \\    
0 & -1
\end{pmatrix}.
\end{align}
This constructing yields the spin-wave energies for any spiral structure in the rotating frame, which may be unfolded to the magnetic unit cell for commensurate structures. It is straightforward to generalize this to more than a single atom in the unit cell \cite{Toth2015}.

The magnon dispersion obtained from this approach is based on linear spinwave theory, which relies on the neglect of bosonic interaction terms arising in the Holstein-Primakoff expansion of spin operators. This is expected to be a reliable approximation at low temperatures where thermally excited magnons are absent and magnon interactions may be neglected. There is, however, one important caveat to this when single-ion anisotropy is included. If one considers an isotropic model with uniaxial anisotropy we may take the easy-axis eigenvalue $\mathcal{A}$  as the only non-vanishing component and in a collinear ferromagnetic model this will yield a spinwave gap of $2AS$. Nevertheless, it is clear that a 2D spin-$1/2$ system cannot order as a consequence of single-ion anisotropy alone and the correct expression is $A(2S-1)$, which only emerges upon normal ordering of the fourth order bosonic terms in the Holstein-Primakoff expansion. Moreover, thermal effects (and the Néel temperature in particular) are strongly affected by magnon interactions and requires a mean-field treatment of magnon interactions as a bare minimum.

Classical Monte Carlo simulations appears to have gained popularity in evaluating critical temperatures in 2D magnets. While the classical treatment is probably somewhat justified for metals, it is by no means clear that it comprises a good approximation for insulators. In fact, standard Weiss mean field theory yields a ratio of $(S+1)/S$ between the critical temperatures of the quantum and classical models. That is, the critical temperature is underestimated by a factor of three for spin $1/2$ and a factor of two for spin-$1$ systems. Alternatively, one may apply a mean-field treatment of fourth order bosonic terms in the Holstein-Primakoff expansion or perform a Tyablikov decoupling of the transverse susceptibility (Green function) \cite{tyablikov1959}. The latter approach is known as the random phase approximation (RPA) and is expected to work well for low temperatures, but probably not in the vicinity of the critical temperature. Nevertheless, it has been applied extensively to estimate Curie temperatures of ferromagnets with exchange constants obtained from first principles \cite{Kudrnovsky2001, Lichtenstein2001, Durhuus2022}. In the case of uniaxial single-ion anisotropy, the RPA (like linear spinwave theory) predicts a spinwave gap of $2AS$ rather than $A(2S-1)$, which obviously represents a severe shortcoming for 2D materials. This can be remedied by an alternative decoupling approach introduced by Callen \cite{callen1963} although that approach is expected to become inaccurate at low spin \cite{FROBRICH2006}. To our knowledge, none of these methods have been benchmarked properly in the presence of anisotropy and we believe that most theoretical estimates of critical temperatures in 2D compounds contain rather severe uncertainties.

\subsection{First principles approaches}
\subsubsection{Ground state}
The most important step towards predicting new magnetic 2D materials is determining the ground state magnetic order. This is a non-trivial task because of the existence of infinitely many possible magnetic ground states. DFT will often converge towards a stationary state that is consistent with the initial guess for a spin configuration and may comprise a local (but not a global) minimum for the total energy. However, even if the spin configuration corresponding to the global minimum in a particular unit cell is found, it may be possible to lower the energy by considering super cells that allow for different spin configurations with lower energy. The zigzag order in FePS$_3$, NiPS$_3$ and RuCl$_3$ are examples of magnetic order that cannot be represented in the minimal unit cell. If the order is not known {\it a priori} a thorough analysis based on super cells thus becomes a daunting task and may require rather extensive super cells if the true ground state has spiral order. A more systematic treatment can be accomplished by applying the generalized Bloch theorem \cite{Sandratskii_1986,Sandratskii1991,Bylander1998,Zimmermann2019}, which allows one to stick with the chemical unit cell in the computations and encode the spiral order through the boundary conditions. The Kohn-Sham Hamiltonian is then parameterized by an ordering vector $\mathbf{q}$ as well as the Bloch momentum $\mathbf{k}$ and one determines the ground state of the Hamiltonian
\begin{align}\label{eq:spiral}
    H_\mathbf{q,k}^\mathrm{KS}&=e^{-i\mathbf{k}\cdot\mathbf{r}}
U_\mathbf{q}(\mathbf{r})H^\mathrm{KS}U_\mathbf{q}^\dag(\mathbf{r})e^{i\mathbf{k}\cdot\mathbf{r}},
\end{align}
where
\begin{align}\label{eq:U}
    U_\mathbf{q}(\mathbf{r})=
\begin{pmatrix}
e^{i\mathbf{q}\cdot\mathbf{r}/2} & 0\\
0 & e^{-i\mathbf{q}\cdot\mathbf{r}/2}.
\end{pmatrix}
\end{align}
Doing this for different ordering vectors yields the ground state order at the vector $\mathbf{q}=\mathbf{Q}$ that minimizes the energy. Determining the ground state thus requires an ensemble of DFT calculations with different order parameters that sample the Brillouin zone. But one avoids having to compare a multitude of spin configurations in extensive super cells. More importantly, the method constitutes a systematic way of obtaining the ground state within the spiral ansatz - even in cases where the spiral order is incommensurate with the unit cell. A major disadvantage of this approach is that it is incompatible with SOC - but since the order is typically determined by bare exchange interactions, it is often assumed that SOC may be included non-selfconsistently once a scalar-relativistic spiral ground state has been obtained \cite{Heide2009, Olsen2016a, Sødequist2023}. FePS$_3$, however, constitutes an important exception to this where selfconsistent SOC is crucial for obtaining a large orbital magnetization and magnetic anisotropy \cite{Kim2021}.

\subsubsection{Exchange constants}
Knowing the ground state order allows one to evaluate the Heisenberg exchange parameters. These may be obtained from first principles methods by simply calculating the energy of various spin configurations and then compare them with the model \cite{Xiang2013, Torelli2018, Olsen2021,Kim2021, Torelli2020}. In the presence of long range interactions, this can become a tedious task and for metals, in particular, it may be difficult to converge a calculation towards a desired spin configuration required for comparison with the model. In addition, for magnons the spin deviates infinitesimally from the ground state and configurations where a single spin has been flipped may alter the electronic structure too much to yield reliable information on the low energy physics contained in the Heisenberg model.

A more systematic treatment can be obtained within DFT using the magnetic force theorem \cite{Liechtenstein1987}, which states that the first order correction to the ground state energy associated with spin deviations is given by the change in Kohn-Sham eigenenergies. This allows for a perturbative treatment and the exchange interactions may be obtained directly from the linear transverse Kohn-Sham susceptibility evaluated on the ground state. For a collinear system the microscopic expression is given by \cite{Durhuus2022}
\begin{align}
J(\mathbf{r}, \mathbf{r}')&=-2B^\mathrm{xc}(\mathbf{r})\chi^{+-}_\mathrm{KS}(\mathbf{r},\mathbf{r}')B^\mathrm{xc}(\mathbf{r}')%\\
%&=\frac{2}{\pi}\int_{-\infty}^{\varepsilon_\mathrm{F}}d\varepsilon B^\mathrm{xc}(\mathbf{r})G^\uparrow_\mathrm{KS}(\mathbf{r},\mathbf{r}', \varepsilon)G^\downarrow_\mathrm{KS}(\mathbf{r}',\mathbf{r}, \varepsilon)B^\mathrm{xc}(\mathbf{r}')\notag
\end{align}
where $\chi^{+-}_\mathrm{KS}$ is the static transverse Kohn-Sham susceptibility and $B^\mathrm{xc}$ is the exchange-correlation magnetic field. The method is readily generalized to the non-isotropic case by transforming to a co-rotating frame similar to the spinwave theory discussed above. It is, however, only possible to extract the locally transverse components of the exchange matrix by this approach, and the longitudinal components (along the spin) has to be extracted by additional calculations of the susceptibility obtained from a spin state constrained to be orthogonal on the ground state magnetization. Finally, once the microscopic components have been extracted these can be mapped to a site-based model (such as the Heisenberg Hamiltonian) by integrating over the relevant magnetic sites. Thus
\begin{align}\label{eq:J_site}
J_{aibj}=\int_{\Omega_i^a}d\mathbf{r}\int_{\Omega_j^b}d\mathbf{r}'J(\mathbf{r},\mathbf{r}'),
\end{align}
where $\Omega_i^a$ is a volume defining magnetic site $a$ in unit cell $i$. The sites may typically be chosen as the magnetic atoms, but the methodology always involves an arbitrary definition of the sites and may also be used to extract magnetic interactions with ligands. The fact that magnetic sites cannot be defined uniquely reflects the limitations of mapping first principles calculations to a site-based model, but it has been shown that magnon energies are typically insensitive to the choice of magnetic sites as long as these contain the majority of the local moments \cite{Durhuus2022}. Alternatively, the approach can be based on Wannier functions centered on the magnetic sites and in that case the exchange couplings in a site basis are extracted directly \cite{He2021}. The ambiguity in defining magnetic sites in Eq. \eqref{eq:J_site} are then reflected in the gauge-dependence of the Wannier functions.

\subsubsection{Magnetic properties of NiBr$_2$}
\begin{figure*}[tb]
    \centering
    \includegraphics[width=\linewidth]{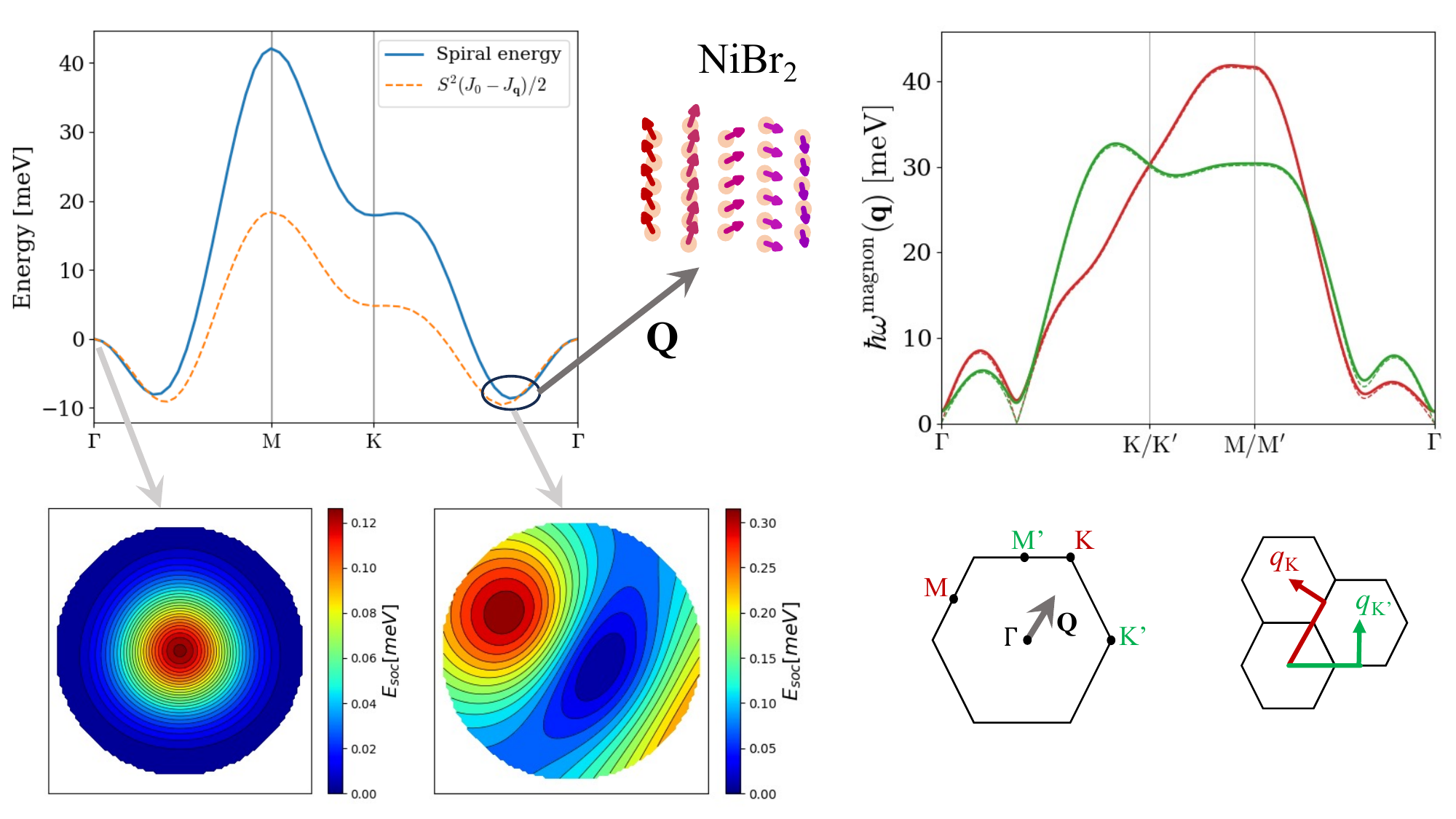}
    \caption{Magnetic properties of NiBr$_2$ calculated with DFT. Upper left: spin spiral energy obtained from the generalized Bloch theorem and from the classical Heisenberg model with isotropic exchange parameters obtained from MFT ($S^2(J_\mathbf{0}$-$J_\mathbf{q})/2$). Lower left: Magnetic anisotropy energies shown as stereographic projections of the upper hemisphere onto the atomic plane. For the ferromagnetic state ($\Gamma$) the direction corresponds to the direction of magnetic moments, whereas the ground state anisotropy energy (at $\mathbf{Q}$)  is with respect to the normal vector of the spiral plane. Right: spin wave dispersion along two paths that would be equivalent based on the space group, but differ due to the broken symmetry induced by $\mathbf{Q}$. Magnon dispersion without SOC is shown as dashed line and exhibits degeneracies at $\Gamma$ and $\mathbf{Q}$.}
    \label{fig:NiBr2}
\end{figure*}
In figure \ref{fig:NiBr2} we illustrate the first principles methods discussed above exemplified by monolayer NiBr$_2$. This material is known as an antiferromagnetic van der Waals bonded material with a Néel temperature of 23 K. The individual layers exhibit spiral order with $\mathbf{Q}=(x, x)$ where $x$ changes continuously from 0.03 to 0.09 between 4.2 and 22.8 K \cite{ADAM19801}. 

We have performed spin spiral calculations using Eq. \eqref{eq:spiral} for $\mathbf{q}$ on the $\Gamma$MK$\Gamma$ path and find a minimum at $\mathbf{Q}=(0.12, 0.12)$. This is in good agreement with the experimental value although we find the magnitude of $\mathbf{Q}$ to be slightly larger. This could originate from either inaccuracies of the LDA functional or from physical differences between bulk and the monolayer. The band width of the spiral energy is roughly 50 meV and provides an estimate of the magnetic exchange energy. We have also calculated the Fourier transform $J_\mathbf{q}$ of the isotropic exchange interactions \eqref{eq:J_site} based on the ferromagnetic structure. For a ferromagnetic ground state, the magnon dispersion would be given by $S(J_\mathbf{0}-J_\mathbf{q})$ and the classical spiral energy can be expressed as $S(J_\mathbf{0}-J_\mathbf{q})/2$. We show this quantity along with the spiral energy and observe that it predicts a minimum in correspondence with the spiral calculation using the generalized Bloch theorem. From the point of view of magnons, this minimum then signifies a dynamical instability of the ferromagnetic state and a recent work has outlined a self-consistent procedure for determining ground state order solely based first principles exchange constants and dynamical instabilities \cite{Tellez-Mora2024}. We emphasize, however, that the exchange constants are evaluated at the ferromagnetic ground state instead of the proper spiral ground state and while the two ground states are expected to yield similar exchange parameters, there will be differences that lead to slightly different predictions. The most reliable value is obtained from the generalized Bloch theorem, which does not involve any assumptions of a site-based model. Similarly, one could argue that the exchange function $J_\mathbf{q}$ may be extracted from the spiral calculations. However, these are most accurately evaluated by infinitesimal perturbation of the spin structure (as encoded by the MFT), but the spiral calculations may involve significant spin rotations at finite $\mathbf{q}$, which renders such an approach inaccurate. Note, however, that the two methods yield the same $J_\mathbf{q}$ when the magnitude of $\mathbf{q}$ is small and the spiral calculations involve small deviations from the ferromagnetic state. In addition to the high symmetry path, we have evaluated $J_\mathbf{q}$ on a regular $12\times12$ mesh and Fourier transformed to get the real space exchange constants. From this we can get all exchange constants that connect sites within 12 unit cells. The ones corresponding to the six shortest distances are $J_1=5.2$, $J_2=-0.3$, $J_3=-4.0$, $J_4=-0.034$, $J_5=-0.097$ and $J_6=0.004$ meV, which is in good agreement with previous calculations \cite{Amoroso2020}

Figure \ref{fig:NiBr2} also shows the magnetic anisotropy energy evaluated in both the ferromagnetic state ($\mathbf{q}=(0, 0)$) and at the ground state ordering vector ($\mathbf{Q}=(0.12, 0.12)$). The ferromagnetic state shows clear easy-plane anisotropy with a very slight trigonal distortion. For the helimagnetic ground state, we display the anisotropy by mapping out the energy as the normal vector to the spiral plane scans the upper hemisphere. We observe a preference of the normal vector to be situated close to the out-of-plane direction (corresponding to an in-plane cycloid), but with a slight tilt in the direction orthogonal to $\mathbf{Q}$. While an in-plane cycloid is expected in an easy-plane monolayer, it is clear that the ordering vector breaks the symmetry rather strongly and the energy is far from invariant under rotations in the plane.

Finally, we show the magnon dispersion calculated according to Eqs. \eqref{eq:dyn_matrix}-\eqref{eq:g} and the associated discussion. We have assumed an in-plane cycloid with easy-plane anisotropy originating from single-ion anisotropy in accordance with the SOC calculations for the ferromagnetic state. For simplicity we have also taken the isotropic exchange constants calculated from the ferromagnetic state. The broken symmetry is reflected by the fact that the dispersion depends on the path relative to the ordering vector. We can for example consider the $\Gamma$K path along $\mathbf{Q}$ (shown in red) and compare with that of a $\Gamma$K path at a $60^\circ$ angle to $\mathbf{Q}$ (shown in green). The single-ion anisotropy yields a spinwave gap of 1.4 meV at $\Gamma$. This is an order of magnitude larger than the SOC energy difference between the hard axis and the easy plane. This is typical for antiferromagnets where the spin wave gap scales with the isotropic exchange parameters as well as the anisotropy itself. Without anisotropy there are massless Goldstone modes (zero energy) at $\Gamma$ as well as the $q$-points that are related by rotational symmetry to $\mathbf{Q}$.

The calculations presented here were performed with the electronic structure package GPAW \cite{Enkovaara2010}, which uses the projector augmented wave method. All calculations were carried out using the LDA exchange-correlation functional and a plane wave basis with a cutoff of 800 eV. The spiral calculations were performed with a $k$-point sampling of $24\times24$ and for the MFT calculations we used a $48\times48$ $k$-point mesh.

\subsubsection{Potential 2D antiferromagnets}
\begin{table}[t]
\begin{tabular}{l|c|c|c|c}
              & Space group & $T_\mathrm{N}$ & Bulk ID & Source   \\
                 \hline
TiCl$_2$                 & $P\bar3m1$ &  85 & 9009121 & \cite{mcguire2017crystal}  \\
TiBr$_2$                 & $P\bar3m1$ &  -  & 26078   & \cite{mcguire2017crystal}  \\
TiI$_2$                  & $P\bar3m1$ &  -  &  -      & \cite{mcguire2017crystal}  \\
TiCl$_2$                 & $P\bar3m1$ &  36 & 1528165 & \cite{mcguire2017crystal}  \\
VTe$_2$                  & $P\bar3m1$ &  -  & 603582  & -  \\
VBr$_2$                  & $P\bar3m1$ &  30 & 246906  & \cite{mcguire2017crystal}  \\
MnCl$_2$                 & $P\bar3m1$ &  2  & 33752   & \cite{mcguire2017crystal}  \\
MnBr$_2$                 & $P\bar3m1$ &  2  & 67500   & \cite{mcguire2017crystal}  \\
MnI$_2$                  & $P\bar3m1$ &  4  & 9009110 & \cite{mcguire2017crystal}  \\
FeO$_2$                  & $P\bar3m1$ &  -  & 9009104 & -  \\
FeI$_2$                  & $P\bar3m1$ &  9  & 9009103 & \cite{mcguire2017crystal}  \\
CoI$_2$                  & $P\bar3m1$ &  11 & 9009100 & \cite{mcguire2017crystal}  \\
NiBr$_2$                 & $P\bar3m1$ &  23 & 9009131 & \cite{mcguire2017crystal}  \\
NiGa$_2$S$_4$            & $P\bar3m1$ &  -  & 634901  & - \\
Mn$_2$Ga$_2$S$_5$        & $P\bar3m1$ &  -  & 634664  & - \\
MnRe$_2$O$_8$            & $P\bar3$   &  -  & 51014   & -  \\
MnGa$_2$S$_4$            & $P3m1$     &  -  & 634670  & - \\
MnAl$_2$S$_4$            & $P3m1$     &  -  & 608511  & - \\
MnIn$_2$Se$_4$           & $P3m1$     &  -  & 639980  & - \\
MnSe                     & $P4/nmm$   &  -  & 162900  & -  \\
CrSe                     & $P4/nmm$   &  -  & 162899  & -  \\
%MnGeBa                   & $P4/nmm$   &  -  & 1539729 & \cite{MnGeBa} \\
VMoF$_5$                 & $P4/n$     &  -  & 1535988 & \cite{VMoO5}\\
FeO$_2$                  & $Pmmn$     &  -  & 9009154 & -  \\
VCl$_2$O                 & $Pmm2$     &  -  & 24380   & - \\
VBr$_2$O                 & $Pmm2$     &  -  & 24381   & - \\
NiC$_4$H$_8$N$_2$O$_4$   & $Pmm2$     &  -  & 4509073 & \cite{mofs} \\
CoCa$_2$O$_3$            & $Amm2$     &  -  & 1531759 & \cite{CoCa2O3} \\
MoCl$_3$                 & $C2/m$     &  -  & 1538556 & \cite{PhysRevMaterials.1.064001}  \\
%CuO$_2$Li$_2$            & $C2/m$     &  -  & 174134  & -  \\
AgF$_2$                  & $P2_1/c$   &  -  & 1509321 & \cite{AgF2} \\
VF$_4$                   & $P2_1/c$   &  -  & 1539645 & \cite{VF4} \\
RuF$_4$                  & $P2_1/c$   &  -  & 165398  & - \\
CrAgP$_2$S$_6$           & $P2/c$     &  -  & 1000192 & \cite{CrAgP2S6} \\
MnClSbS$_2$              & $P2/m$     &  -  & 151925  & -  \\
MnBrSbS$_2$              & $P2/m$     &  -  & 1528449 & \cite{MnBrSbS2} \\
MnBrSbSe$_2$             & $P2/m$     &  -  & 1528451 & \cite{MnBrSbS2} \\
MnISbSe$_2$              & $P2/m$     &  -  & 2013470 & \cite{MnISbSe2} \\
CoBr$_2$Sb$_2$O$_3$      & $P2_1/m$   &  -  & 418858  & - \\
VSeO$_4$                 & $P\bar1$   &  -  & 69994   & - \\
MnSb$_2$F$_{12}$         & $P\bar1$   &  -  & 1535152 & \cite{MnSb2F12}
\end{tabular}
\caption{List of antiferromagnetic monolayers (with PBE) that have known parent bulk van der Waals bonded compounds.}
\label{tab:predictions}
\end{table}
Numerous suggestions for novel 2D magnetic materials have appeared in the literature over the recent years. As such, the C2DB alone contains 1682 magnetic 2D materials, with 606 compounds that are both dynamically stable and situated in close vicinity to the convex hull. These predictions range from unconstrained bottom-up design to exploring the possibility of {\it in silico} exfoliation of bulk van der Waals bonded materials. While we recognize the possibility of bottom synthesis by chemical vapor deposition or similar methods, we have chosen to focus on 2D materials that has a bulk van der Waals bonded parent compound. In Ref. \cite{Mounet2018}, the Inorganic Crystal Structure Database (ICSD) \cite{Allmann2007} and the Crystallography Open Database (COD) \cite{Graulis2011} were screened for possible exfoliable materials, which resulted in 56 predicted stable 2D magnets. A similar approach was applied in Ref. \cite{Torelli2020a}, which yielded 85 ferromagnetic and 61 antiferromagnetic compounds (without considerations of stability). These are all part of the C2DB, which thus serves as a convenient starting point when searching for 2D antiferromagnets. In particular, the stability criterion of the C2DB allow us to exclude materials that become unstable in the 2D limit. Moreover, while all monolayers in the C2DB are represented in ferromagnetic configurations, an effective nearest neighbor exchange constant has been extracted for stable materials and negative values imply that the ferromagnetic order cannot represent the ground state. In Ref. \cite{sødequist2023magnetic} the magnetic ground state was extracted from spin spiral calculations for all stable materials in C2DB containing a single magnetic atom in the unit cell and more than half of these exhibited non-collinear anti-ferromagnetic order while 11{\%} had collinear antiferromagnetic order.

In table \ref{tab:predictions} we list all dynamically stable compounds from the C2DB that have been shown to have an unstable ferromagnetic ground state by direct evaluation of the effective nearest neighbor interaction \cite{Torelli2020a}. Only materials that has an experimentally characterized bulk van der Waals bonded parent compound are shown. We state the non-magnetic space group of the monolayer as well as the bulk Néel temperature and database identifier of the parent bulk structure from the ICSD ($<10^6$) or the COD ( $>10^6$). The bulk transition metal dihalides were discussed in great detail in Ref. \cite{mcguire2017crystal} and the ground state of the corresponding monolayers were extracted by LDA spin spiral calculations in Ref. \cite{Sødequist2023}. Some of the remaining materials have also been investigated in detail by first principles methods. For example the VX$_2$O (X=Cl, Br), which have been shown to be ferroelectric \cite{VOX2_multiferroics} and AgF$_2$, VF$_4$ and RuF$_4$, which exhibits altermagnetism \cite{sødequist2024twodimensional}. The bulk parent compounds of the materials in table \ref{tab:predictions} have been characterized experimentally with respect to crystal structure, but the magnetic order of the majority of these is largely unexplored. We believe that most of the compounds could comprise potentially exfoliable 2D antiferromagnets and it would be highly desirable (and motivational for experimental exfoliation) if the magnetic properties were analyzed in detail by first principles methods prior to experimental scrutiny. Finally we note that several monolayers in the C2DB could have antiferromagnetic ground states despite having ferromagnetic effective nearest neighbor exchange constants. This is often the case for compounds having spiral ground states like for example NiBr$_2$ discussed above. Table \ref{tab:predictions} thus represents a minimal compilation of exfoliable monolayers from the C2DB where the ground state is predicted to be antiferromagnetic from DFT (using the PBE functional).

\section{Outlook and challenges}
We have provided a summary of experimentally known 2D anti-ferromagnets and outlined the computational approaches that may be applied to unravel magnetic properties from first principles calculations. Such calculations are, however, accompanied by significant challenges that were not addressed. The most serious issue is the fact that antiferromagnetism is typically associated with strong correlation and the accuracy of DFT thus becomes questionable. For most known magnetic materials DFT seems to predict the correct magnetic order but estimates of the exchange constants may be rather dependent on the applied exchange-correlation functional. Typically, Hubbard corrections are included to remedy the localization errors associated with semi-local functionals, but the magnitude of exchange constant may be sensitive to the applied value of U. For example, the experimental magnon dispersion MnPS$_3$ can be accurately reproduced within an energy mapping approach using PBE+U with U=2eV (arbitrarily chosen), but the magnons band width varies by a factor of 2.5 when U is varied between 1 and 5 eV \cite{Olsen2021}. While a unique value of U may be extracted from self-consistent calculations, it remains to be seen whether such approaches yields values that reproduce the experimental magnon spectrum. In addition, insulators like FePS$_3$ and  RuCl$_3$ are predicted to be gapless without U and Hubbard corrections thus seems to be a crucial ingredient for simulations of such materials.

At a more fundamental level it is far from clear how magnetic order evolves when bulk van der Waals bonded materials are thinned down to the monolayer limit. Referring to table \ref{tab:antiferromagnets} one observes that for some compounds (for example NiI$_2$ and MnPSe$_2$) the transition to the monolayer limit is associated with significant reduction of Néel temperatures, while for others the Néel temperature is largely unaffected. The most obvious explanation would be that for some materials the magnetic anisotropy plays a prominent role whereas for others the interlayer coupling leads to increased magnetic stability in the bulk compounds. But the cases of MnPS$_3$ and MnPSe$_3$ seem to defy such an interpretation. MnPS$_3$ is nearly isotropic with a very weak easy axis and the Néel temperature is largely unaffected by the transition to the monolayer limit. On the other hand, MnPSe$_3$ has a rather strong easy plane and would not be expected to order by the Mermin-Wagner theorem (rotational symmetry is conserved by the Néel type order in this material). Yet the magnetic order is definitely conserved albeit with a significant reduction of Néel temperature. In general, materials with strong easy planes are expected to exhibit a Kosterlitz-Thouless transition below which quasi long range order is maintained due to algebraic decay of spin correlations. Such decay is only well defined in the limit where spin is strictly confined to a plane, which is never the case in real materials. Nevertheless, Kosterlitz-Thouless physics could play a prominent role in such systems and it may be hard to distinguish the mechanisms responsible for maintaining magnetic order experimentally. There is thus a strong need for thorough theoretical (or numerical) analysis of how the Kosterlitz-Thouless physics plays out in the case of finite (but not infinite) anisotropy. These considerations are equally valid for ferromagnets, which may constitute a better starting point for such analysis. In particular, the case of monolayer CrCl$_3$ is ferromagnetic with an easy-plane and has been shown to order below 13 K with a critical exponent of $\beta=0.227$, which agrees well with a Kosterlitz-Thouless transition of the ideal $XY$-model \cite{Bedoya-Pinto2021}. The concept of critical behavior and universality classes is thus a highly useful tool for unraveling the nature of magnetic transition, but more experimental as well as theoretical work is definitely needed on this. It is, for example, not clear why (quasi) long range order seems to be absent in the case of 2D ferromagnetic CrGeTe$_3$ \cite{Gong2017a}, which has a similar easy plane.

On the pure theoretical side, the development of accurate methods for predicting critical temperatures is pertinent. This cannot be accomplished directly with methods such as DFT and requires mapping to site-based (Heisenberg) models, but even in that case it is not clear how to proceed for quantitative estimates. Quantum Monte Carlo methods and renormalization group theory is presently not at a point where they may be routinely applied to extract critical temperatures in real materials and most groups resort to classical MC simulations for extracting thermal properties with {\it ab initio} exchange constants. The fact that Weiss mean field theory severely overestimates critical temperatures in the classical limit is, however, a strong indication that similar systematic errors are associated with classical MC simulations. A quantum mechanical analysis can be carried out with RPA or renormalized spin wave theory, but the accuracy of such approaches is currently not clear for ferromagnets and it is likely that the performance becomes even more dubious for antiferromagnets where the ground state is not well described by the classical configuration that minimizes the energy. For bulk materials, it is often possible to compare calculated exchange parameters to the magnon dispersion extracted from inelastic neutron scattering. But for 2D materials, this is typically not possible and the critical temperatures constitute one of the only observables that may be used to validate computational approaches. In our view, the most important step required for progress on the computational side, is the development (and benchmarking) of methods to extract reliable thermal properties - possibly based on bulk materials where the exchange constants are often accurately known from experiments. Only then will it be possible to benchmark computationally methods for extracting exchange constants and we will be a large step closer to a reliable computational framework with predictive power.

\newpage
\bibliography{bibliography.bib}

\end{document}